\documentclass[seceq]{ptptex}

\newcommand{\la}{\langle}
\newcommand{\ra}{\rangle}

\newcommand{\beq}{\begin{eqnarray}}
\newcommand{\eeq}{\end{eqnarray}}
\newcommand{\bfl}{\begin{flushleft}}
\newcommand{\efl}{\end{flushleft}}

\notypesetlogo                       
\title{
Radiative Decays of $q\bar{q}$ Chiral States in the $\widetilde{U}(12)$-Scheme
}

\author{
Tomohito \textsc{Maeda}$^1$, 
Kenji \textsc{Yamada}$^1$, 
Masuho \textsc{Oda}$^2$, 
and Shin \textsc{Ishida}$^3$
}

\inst{
$^{1}$Department of Engineering Science, 
Junior College Funabashi Campus, \\
Nihon University, Funabashi 274-8501, Japan\\
$^{2}$Faculty of Engineering, Kokushikan University, 
Tokyo 154-8515, Japan\\
$^{3}$Research Institute of Science and Technology, 
College of Science and Technology, \\
Nihon University, Tokyo 101-8308, Japan
}

\abst{
The radiative transitions between ground-states (GS) 
of light $q\bar{q}$ mesons 
are investigated in the 
$\widetilde{U}(12)$-scheme. 
In this scheme the rich decay-spectra are 
offered even in transition between GS due to 
the appearance of chiral states. 
As a result the radiative decay widths of the ordinary 
$V\to P \gamma$ process are well reproduced, and 
furthermore, the predicted width 
of $b_{1}(1235) \to \pi \gamma$ process ($\Gamma_{\rm theor}=229$ keV), 
by assigning $b_{1}(1235)$ as 
a member of the ${}^{3}S_{1}$ chiral states $A^{(E)}$, is 
in good agreement with experimental one ($\Gamma_{\rm exp} =230 \pm 60 $ keV). 
From this results it is indicated 
that $b_{1}(1235)$ meson, classified as ${}^{1}P_{1}$ state 
in conventional non-relativistic (NR) classification scheme, 
is a good candidate of chiral state 
in the $\widetilde{U}(12)$-scheme. 
The other radiative transition widths of some chiral states 
are also predicted. 
Accordingly the predicted values 
given here, will provide useful insights in their search at BES.
}
\begin{document}
\maketitle
\section{Introduction}
\bfl
\ \ ({\it $\widetilde{U}(12)$-scheme and $q\bar{q}$ chiral states}) \\ 
\efl

 The $\widetilde{U}(12)$-scheme, which has been proposed several years ago, 
is a Lorentz covariant framework for describing 
the composite hadron system based on ``static'' ${U}(12)_{SF}$-symmetry, 
embedded in $\widetilde{U}_{SF}(12) \otimes O(3,1)_{L} $-space.\footnote{The word 
``static'' implies that 
the symmetry is imposed only at the rest frame of hadrons. 
For more detail, see Ref.~\citen{Dshep}. 
} 
The wave function (WF) of composite hadron 
is generally described as irreducible tensors 
of $U(12)_{SF}$-group 
at their rest frame. 
The ${U}(12)_{SF}$-group includes, 
in addition to the conventional non-relativistic $SU(6)_{SF}$-group, 
the new symmetry $SU(2)_{\rho}$ 
which is naturally introduced in connection with the covariant treatment 
of the constituent confined quarks.{\footnote{{The new degree of 
freedom corresponding to the $SU(2)_{\rho}$-symmetry is called 
the $\rho$-spin, after the well-known $\rho\otimes\sigma$-decomposition of 
Dirac matrices. }
} 
By inclusion of this extra $SU(2)$ spin freedom, 
it leads to the possible existence of many new states, 
called chiral states or chiralons.\cite{u-12,Cov} 
Chiral states are described with at least one Dirac spinors with negative 
$\rho_{3}$ (the third component of the $\rho$-spin)-
eigen-value, 
and never appear in the non-relativistic quark model (NRQM). 
On the other hand, the conventional states appearing in 
NRQM (called Pauli states or Paulons) 
are described with Dirac spinors with  all positive $\rho_{3}$-eigen-values.

In the $\widetilde{U}(12)$-scheme the GS of light $q\bar{q}$ meson system 
is assigned as ${\bf 12} \times {\bf {12}^{*}}={\bf 144}$-representation 
of the static $U(12)_{SF}$ group and 
they are classified into the eight 
flavor nonets with 
$J^{PC}$ (see, Table \ref{tab:JPC}). 
As is shown in this table, the ${\bf 144}$-plet contains
two pseudo-scalar (labeled $P_{s}^{(N)}$ and $P_{s}^{(E)}$) nonets, 
two vector (labeled $V_{\mu}^{(N)}$ and $V_{\mu}^{(E)}$) nonets, 
two scalar (labeled $S^{(N)}$ and $S^{(E)}$) nonets, 
and two axial-vector 
(labeled $A_{\mu}^{(N)}$ and $A_{\mu}^{(E)}$) nonets.{\footnote{All these states 
are chiral state by the implications above mentioned. 
However, as explained in the next section, 
only vector $\rho$-meson nonets becomes pure Pauli state, 
being a linear combination of the $V_{\mu}^{(N)}$ and $V_{\mu}^{(E)}$-states.
}} 
Here we emphasize that, in the $\widetilde{U}(12)$-scheme, 
the scalar $0^{++}$ and axial-vector $1^{++/+-}$ nonets 
appear not only 
in excited P-wave state but also in the S-wave GS. 
These GS chiralons are expected to have the lower mass than 
the $P$-wave states. 
The $\sigma$-nonet are regarded as an appropriate candidate 
of the $0^{++}$ GS chiralon,{\cite{Cov}} 
while it will be shown that $b_{1}(1235)$ 
meson is also another good candidates of $1^{+-}$ GS 
chiralon through this analysis. 
On the other hand, it is shown that 
experimentally well known $a_{1}(1260)$-meson, decaying with partial width 
$\Gamma_{\rm exp} (a_{1}(1260)\to \pi \gamma)=640\pm 240$ keV,{\cite{PDG2004}} 
is not a pure GS chiralon.
\bfl
\ \ ({\it The important features of our scheme and radiative decays})\\
\efl

In the radiative transition between light-quark meson system 
the final mesons generally move relativistically. 
In addition to this, considering that all actual physical observations 
are made through not quarks but hadrons, 
covariant treatment for the center of mass (CM) motion of hadron 
is absolutely necessary. Although the NRQM may incorporate 
the effect of relativistic motion of the constituent quarks, 
but it is unable to give the conserved current concerning the composite hadron. 
On the other hand, the $\widetilde{U}(12)$-scheme has remarkable features 
that hadrons are treated in manifestly 
covariant way and the conserved effective hadron currents 
are explicitly 
given in terms of hadron variable themselves. 

\bfl
\ \ ({\it Radiative decays and our previous work})\\ 
\efl

Here it will be worth mentioning the relation between the $\widetilde{U}(12)$-scheme and 
our previous scheme for the relevant problem.
Before possible existence of chiralons being noticed, 
we had been investigated systematically 
radiative decays of light through heavy quark meson systems 
in framework of the covariant oscillator 
quark model{\cite{oldCOQM}} (COQM).{\footnote{The COQM has a long history to development.
In Ref.~{\citen{MassCOQM}} it is applied to investigate the light-quark meson 
spectra, including the one-gluon-exchange effect, with considerable success. 
}} 
As the results, it had been shown that qualitative 
features of experimental data are well reproduced except for some cases. 
The COQM is also based on the $\widetilde{U}(12)\otimes O(3,1)$-scheme, but with static 
$SU(6)_{SF}$-symmetry. Thus, the coverage of the COQM is limited to the reaction 
of Pauli states. 
The new $\widetilde{U}(12)$-scheme with static $U(12)_{SF}$-symmetry 
is extended, keeping the above mentioned covariant properties, 
to be able to describe the chiral states 
in addition to Pauli states. 

\bfl
\ \ ({\it Purpose of this work}) 
\efl

In this report, 
now taking into account the existence of chiral states, 
we shall examine the radiative transitions 
among GS mesons in the $\widetilde{U}(12)$-scheme, 
leading to some useful guidance for 
searching the chiral particle in BES experiment. 
\section{Ground State $q\bar{q}$ Chiral Mesons and their Wave Functions}
\bfl
({\it The $q\bar{q}$-meson wave functions in $\widetilde{U}(12)$-scheme})
\efl

First we recapitulate briefly the framework of $\widetilde{U}(12)$-scheme so far 
as required for the relevant application. 
In our scheme, mesons are unifiedly described by the bi-local 
Klein-Gordon field with one each of lower and upper indices 
in the boosted $U(12)_{SF} \otimes O(3)_{L}$ 
space{\footnote{For simplicity, only expressions for the GS are shown 
in the following.}} as 
\beq
\Phi(x_{q},x_{\bar{q}})_{A}{}^{B}=\int \frac{d^{3} 
{\bf P}}{\sqrt{(2\pi)^{3} 2P_{0}}} 
( e^{+iPX} \Phi(x, P)^{(+)}_{A}{}^{B} + e^{-iPX} \Phi(x, P)^{(-)}_{A}{}^{B} ), 
\label{eq:a}
\eeq
where $A=(\alpha,a) ( B=(\beta,b)) $ denotes Dirac spinor and flavor indices
respectively, $x_{q}$ ($x_{\bar{q}}$) represents a space-time coordinate of 
quark (anti-quark) 
which is related to the CM (relative) coordinate 
for composite meson as 
$X=(m_{q}x_{q}+m_{\bar{q}}x_{\bar{q}})/(m_{q}+m_{\bar{q}})$ ($x=x_{q}-x_{\bar{q}}$) 
($m_{q}$ ($m_{\bar{q}}$) being the quark (anti-quark) mass parameter). 
The ${P_{\mu}}$ denotes CM 4-momentum of the relevant meson. 
The above bi-local field is supposed to 
satisfy the Klein-Gordon type equation 
\beq
\left[ 
\frac{\partial^2}{\partial X^{2}}-{\cal{M}}^2(x,\frac{\partial}{\partial x}) \right] 
\Phi(X,x)_{A}{}^{B}=0, 
\label{KGeq}
\eeq
where ${\cal{M}}^2(x,\frac{\partial}{\partial x})$ is 
the $U(4)$($\supset SU(2)_{\sigma}\otimes SU(2)_{\rho}$)-spin 
independent squared mass operator 
including only central confining-force potential. 
As the basis to expand the above internal wave function 
$\Phi(x,P)_{A}{}^{B}$, we use the ``$LS$-coupling'' type direct products, 
\beq
\Phi (x,P)_{A}^{(\pm)}{}^{B} \approx f(x,P)~W^{(\pm)}_{A}{}^{B}(P),
\label{LS}
\eeq
where the definite-metric type 4-dimensional oscillator function $f(x,P)$ 
and the extended Bargmann-Wigner (BW) 
spinor WF $W^{(\pm)}_{A}{}^{B}(P)$ are taken 
for the respective parts.{\footnote{In the present application, 
we neglect the form factor coming from 
the overlapping of the space-time internal WF, 
since $\int d^{4} x f(x,P^{'})f(x,P)e^{\mp\frac{m_{2,1}}{m_{1}+m_{2}}qx} \approx 1 $ 
is expected for transition between the GS. 
}} 
Decomposing the $W^{(+)}_{A}{}^{B}(P)$ into irreducible definite 
$J^{PC}$-components, 
their detailed forms for relevant meson system are given as{\cite{Cov}} 
\beq
W^{(+)}_{A}{}^{B}(P) 
&=&(1)_{\alpha}^{\beta}(S^{(N)}(P))_{a}^{b} 
+(i\gamma_{5})_{\alpha}^{\beta}(P^{(N)}(P))_{a}^{b}
\nonumber\\
&+&(-v\gamma)_{\alpha}^{\beta}(S^{(E)} (P))_{a}^{b}
+(-\gamma_{5}v\gamma)_{\alpha}^{\beta}(P^{(E)}(P))_{a}^{b}\nonumber\\
&+&(i\tilde{\gamma}_{\mu})_{\alpha}^{\beta}(V_{\mu}^{(N)} (P))_{a}^{b}
+(i\gamma_{5}\tilde{\gamma}_{\mu})_{\alpha}^{\beta}(A^{(N)}_{\mu} (P))_{a}^{b}\nonumber\\
&+&(-i\sigma_{\mu\nu}v_{\nu})_{\alpha}^{\beta}(V_{\mu}^{(E)} (P))_{a}^{b}
+(\gamma_{5}\sigma_{\mu\nu}v_{\nu})_{\alpha}^{\beta}(A_{\mu}^{(E)} (P))_{a}^{b}, 
\label{all}
\eeq 
where the $v_{\mu}={P_{\mu}}/{M}$ denotes 4-velocity of the relevant 
meson.{\footnote{The negative frequency (creation) part of the internal 
spin WF, $W^{(-)}(P)$ can be obtained by substituting the CM momentum 
by $P_{\mu} \to -P_{\mu}$
in the relevant positive frequency (annihilation) part $W^{(+)}(P)$ 
owing to the crossing relation for the meson field. 
}} 
Note that the expansion basis of the BW spin function 
consists of totally 16 members in the $\widetilde{U}(4)$-space.
\begin{table}[htbp]
\caption{Ground-state light $q\bar{q}$ meson nonets 
in $\widetilde{U}(12)$-scheme. Their spin WFs and quantum numbers of the relevant mesons 
are shown. 
The symbol $\chi$ denotes the chirality eigin-value of the relevant meson WF. 
Its definition and physical meaning are discussed in the text. 
}
\begin{center}
\begin{tabular}{lcccccccc}
\hline
\hline
Meson Nonets:&$P^{(N)}$&${P}^{(E)}$&$V^{(N)}$&$V^{(E)}$&$A^{(N)}$&$A^{(E)}$& 
$S^{(N)}$&$S^{(E)}$\\
\hline
$J^{PC\chi}:$&$0^{-+-}$     &$0^{-++}$     &$1^{--+}$ &$1^{---} $
&$1^{+++}   $&$1^{+--}$&$0^{++-}$&$0^{+-+}$\\
\hline
$W(P)^{(+)}$:&$\frac{i\gamma_{5}}{2}$&$\frac{i\gamma_{5} v\gamma}{2}$&
$\frac{i\tilde{\gamma}_{\mu}}{2}$&$\frac{-i\sigma_{\mu\nu} v_{\nu}}{2}$
&$\frac{i\gamma_{5}\tilde{\gamma}_{\mu}}{2}$&
$\frac{\gamma_{5}\sigma_{\mu\nu}v_{\nu}}{2}$&
$\frac{1}{2}$&$-\frac{v\gamma}{2}$\\
\hline
\end{tabular}
\label{tab:JPC}
\end{center}
\end{table}
\bfl
({\it Chiral representation for two kinds of 
Vector / Pseudo-scalar nonets})
\efl

As was explained in the previous section, 
in our scheme, two kinds of vector and pseudo-scalar 
meson nonets appear in GS, respectively. 
To make a discrimination of respective meson, 
it is useful to see the eigen-value of the chirality operator 
of the relevant bi-spinor WF. 
The chirality transformation{\cite{Ishida}} for the bi-spinor WF is defined 
as 
\beq
(\gamma_{5})_{\alpha'}{}^{\alpha} W(P)_{\alpha}{}^{\beta}(-\gamma_{5})_{\beta}{}^{\beta'}
\equiv \chi W(P)_{\alpha'}{}^{\beta'}, 
\label{chi-t}
\eeq
where $\chi$, denotes the eigen value of chirality operator for 
the composite meson system. In case that the relevant composite meson 
consists of quarks with same chiralities (i.e. 
Left$\times$Left or Right$\times$Right combination), 
then the $\chi$ becomes 
$\chi=+1$, and vice versa. 

The $V^{(N)}$-type vector meson WF is given by 
\beq 
W(P)^{(+)}_{A}{}^{B}
&=&(\frac{i}{2} \tilde{\gamma}_{\mu}\epsilon(P)_{\mu} 
V^{(N)}_{a}\frac{\lambda^{a}}{\sqrt{2}}  )_{A}{}^{B} \nonumber \\
&\stackrel{{\bf P}=0}{\rightarrow}&
(\frac{i}{2} 
\gamma_{i} \epsilon(\mbox{\boldmath $P$}=0)_{i} V^{(N)}_{a}\frac{\lambda^{a}}{\sqrt{2}})_{A}{}^{B} 
\nonumber \\ 
&=&\left[ \frac{ \mbox{\boldmath $\sigma$} \cdot \mbox{\boldmath $\epsilon$}(
\mbox{\boldmath $P$}=0)}{\sqrt{2}}V^{(N)}_{a}
\frac{\lambda^{a}}{\sqrt{2}}
\frac{(|+\ra _{\rho} {}_{\bar{\rho}}\la +| + 
 |-\ra _{\rho} {}_{\bar{\rho}}\la -|)}{\sqrt{2}}  \right]_{A}^{B}  
\label{VN}
\eeq
with $\tilde{\gamma}_{\mu} \equiv \gamma_{\mu}+(\gamma\cdot v)v_{\mu}$ satisfying 
$\tilde{\gamma}_{\mu} v_{\mu}=0$, where
$\epsilon_{\mu}$ and $V^{(N)}_{a} ( { a=0 \sim 8 })$ represent the polarization vector and the 
flavor WF of the 
relevant vector meson, respectively. 
In the third expression of Eq.(\ref{VN}), the WF 
is written in $SU(6)_{SF}\otimes SU(2)_{\rho}$ 
form. 
This WF is the eigen-function of chirality operation defined as Eq.(\ref{chi-t}), 
with the eigen-value $\chi=+1$ reflecting its chiral 
$SU(3)_{L}\otimes SU(3)_{R} $-representation, 
$({\bar{1}}_{L},1_{R}) 
\oplus ({1}_{L},{\bar{1}}_{R}) / 
({8}_{L},1_{R}) \oplus ({1}_{L},8_{R})$.{\footnote{This type representation is well 
known and 
the common expression is used to various models. 
See, for example, Ref.~{\citen{Glozman}}. 
Here it should be noted that the definition of chirality transformation 
Eq.(\ref{chi-t}) is deduced from the basic structure of our classification scheme. 
}} 
Here it may be notable that, as shown in the last line of Eq.(\ref{VN}), 
the $V^{(N)}$-WF is a superposition pure-Pauli states and pure-chiral states. 

On the other hand, another type of vector meson WF, $V^{(E)}$ as given by 
\beq 
W(P)^{(+)}_{A}{}^{B}
&=&(\frac{-i}{2} \sigma_{\mu\nu}v_{\nu} \epsilon(P)_{\mu} 
V^{(E)}_{a}\frac{\lambda^{a}}{\sqrt{2}}  )_{A}{}^{B} \nonumber \\
&\stackrel{{\bf P}=0}{\rightarrow}&
(\frac{1}{2} \sigma_{i4} \epsilon(\mbox{\boldmath $P$}=0)_{i} V^{(E)}_{a}\frac{\lambda^{a}}{\sqrt{2}})_{A}{}^{B} 
\nonumber \\ 
&=&\left[ \frac{ \mbox{\boldmath $\sigma$} \cdot \mbox{\boldmath $\epsilon$}(
\mbox{\boldmath $P$}=0)}{\sqrt{2}}V^{(E)}_{a}
\frac{\lambda^{a}}{\sqrt{2}}
\frac{(|+\ra _{\rho} {}_{\bar{\rho}}\la +| - 
 |-\ra _{\rho} {}_{\bar{\rho}}\la -|)}{\sqrt{2}} 
 \right]_{A}^{B}, 
\label{VE}
\eeq
is also contained in the GS of the $\widetilde{U}(12)$-scheme, which is corresponding 
to the representation $(\bar{3}_{L},3_{R}) \oplus (3_{L},\bar{3}_{R})$ of the chiral 
$SU(3)_{L}\otimes SU(3)_{R} $. Thus, the eigen-value of chirality is $\chi=-1$. 
This $V^{(E)}$-WF is the orthogonal superposition (to the $V^{(N)}$-WF) of Pauli- and 
chiral-states.
Here it should be noted that 
generally the spin WF of physical $\rho$-nonets 
is given as a superposition of the $(N)$ and $(E)$ components, 
since they have the same 
$J^{PC}$. 
Being based on its success{\cite{Oda}} with $SU(6)_{SF}$-description 
for $\rho$-nonet, it seems that its WF 
should be taken as the same form as in our previous 
COQM, containing only Pauli-states. 
This is obtained by taking the following two linear combinations of $(N)$ and $(E)$, 
$V$ and $V^{'}$ with equal weight, 
\beq
V \equiv (V^{(N)} + V^{(E)})/{\sqrt{2}} , \ \ \ 
V^{'} \equiv (V^{(N)} - V^{(E)})/{\sqrt{2}} \ .
\label{Vec}
\eeq
Here the $V$ represents $\rho$-nonet as pure Pauli state, 
while the $V^{\prime}$ represents the other nonets as ``pure chiral state'', which 
is newly appeared in $\widetilde{U}(12)$-scheme. 
They are mutually orthogonal tensors in static $SU(2)_{\rho}$-space. 
The resultant WF of physical $\rho$-nonet is described as 
\beq 
W(P)^{(+)}_{A}{}^{B}
&=&(\frac{1}{2\sqrt{2}} i\tilde{\gamma}_{\mu}(1+iv\gamma)\epsilon(P)_{\mu} 
V_{a}\frac{\lambda^{a}}{\sqrt{2}} )_{A}{}^{B} \nonumber \\
&\stackrel{{\bf P}=0}{\rightarrow}&
\left[ \frac{ \mbox{\boldmath $\sigma$} \cdot \mbox{\boldmath $\epsilon$}(
\mbox{\boldmath $P$}=0)}{\sqrt{2}}V^{(N)}_{a}
\frac{\lambda^{a}}{\sqrt{2}} 
(|+ \ra _{\rho} {}_{\bar{\rho}} \la + |  ) \right]_{A}^{B} .
\label{V}
\eeq
It should be noted that the above WF no longer has definite chirality, and 
instead it includes only the pure eigen-states with 
$(\rho_{3}, \bar{\rho_{3}})=(+1,+1)$. 

The above statement also holds for the two-type pseudo-scalar WFs. 
However, in contrast to the case of vectors, 
the WF of 
$\pi$-nonet, which is denoted by $P$, should 
belong to the same chiral representation as $\sigma$-nonet 
from the consideration on linear realization of chiral symmetry. 
The resultant WF of $\pi$-nonet is described as 
\beq
W(P)^{(+)}_{A}{}^{B}
&=&(\frac{1}{2} i{\gamma}_{5} 
P_{a}\frac{\lambda^{a}}{\sqrt{2}} )_{A}{}^{B}, \ \ 
 P \equiv P^{(N)} . 
\label{P}
\eeq 
\bfl
({\it Effective Meson Currents })
\efl

In order to treat the interaction of $q\bar{q}$-mesons 
with the electro-magnetic (EM) 
field $A_{\mu}(X)$, 
we start from the free action of the meson system, 
\beq
{S}_{0}&=&\int d^4 x_{q} ~ d^4 x_{\bar{q}} 
\la {\cal L} _{0}(x_{q},x_{\bar{q}}) \ra_{S,F} =\int d^4 x ~ d^4 X \la {\cal L} _{0}(x,X) \ra_{S,F}
\equiv \int d^4 X  {\cal L} _{0}(X), \nonumber\\ 
&&{\cal L} _{0}(X) = \int d^{4} x \la 
\bar{\Phi}
(x, X) \left( \frac{\partial^2}{\partial {X}_{\mu}^2} 
- {\cal{M}}(x)^{2} \right) \Phi (x, X) \ra_{S,F} \ .
\label{aiu}
\eeq 
Here, $\la \cdots \ra_{S,F}$ denotes taking the 
trace over Dirac spinor and flavor indices. 
The Pauli-conjugate of the bi-spinor function is defined as 
$\bar{\Phi}_{\alpha}{}^{\beta}\equiv (\gamma_{4})_{\alpha}{}^{\alpha^{'}}
({\Phi^{\dagger}})_{\alpha^{'}}{}^{\beta^{'}} (\gamma_{4})_{\beta^{'}}{}^{\beta}$. 
Note that the action ({\ref{aiu}}), which 
is the simplest form to lead to our basic Eq.({\ref{KGeq}}), 
has the $\widetilde{U}(12)_{SF}$-symmetry. 
Then it is necessary to implement 
the following modification. 
We shall explain this procedure 
by citing a concrete descriptions in the following : 
The action ({\ref{aiu}}) yields the amplitude in momentum representation, 
\beq
{\tilde{A}}_{0} &=& \frac{2\pi}{\sqrt{4P_{0}P_{0}^{'}}} \delta^{4}(P-P^{'}) 
\left( P_{\mu}^{2}-M_{0}^{2} \right)
\{
\la {\bar{W}_{H}^{(-)}} (P^{'})  W_{H}^{(+)}(P) \ra_{S} \nonumber\\
&&~~~~~~~~~~~~~~~~~~~~~~~~~~~~~~~~~~~~~~~
+\la {\bar{W}_{\bar{H}}^{(+)}} (P^{'})  W_{\bar{H}}^{(-)}(P) \ra_{S}
\}.
\label{tildeinv}
\eeq
In Eq.(\ref{tildeinv}), $\la \cdots \ra_{S}$ represents taking 
the trace over the only Dirac spinor 
indices, while the flavor indices are specified here. 
The first (second) term of Eq.(\ref{tildeinv}) 
describes annihilation of 
the meson $H$ (anti-meson $\bar{H}$) with momentum $P$ 
and creation of the meson $H$ (anti-meson $\bar{H}$) with momentum $P^{'}$. 
In order to guarantee the static $U(12)_{SF}$-invariance of the 
amplitude (\ref{tildeinv}) embedded in the 
$\widetilde{U}(12)_{SF}$-space, we 
introduce the vertex 
factor, 
\beq
F_{U}(v) \equiv -iv \cdot \gamma, 
\eeq
called unitarizer.{\cite{Ishida}} 
Inserting the unitarizer in appropriate 
places of the above amplitude 
makes it static $U_{SF}(12)$-invariant and leads to 
the correct sign of mass term 
for the chiral states as well as the Pauli states. 
By using this prescription, 
Eq.({\ref{tildeinv}}) is replaced by 
\beq
{\tilde{A}}_{0} \to A_{0} &=&
\frac{2\pi}{\sqrt{4P_{0}P_{0}^{'}}} \delta^{4}(P-P^{'}) 
\left( P_{\mu}^{2}-M_{0}^{2} \right)
\{ 
\la { \bar{W}_{H}^{(-)}(P^{'})} F_{U}(v^{'})
 W_{H}^{(+)}(P)  F_{U}(v^{}) \ra_{S}
\nonumber\\
&&~+ \la { \bar{W}_{\bar{H}}^{(+)}}(P^{'}) 
 F_{U}(v^{'}) W_{\bar{H}}^{(-)}(P) F_{U}(v^{}) \ra_{S} \}. 
\label{freeKG}
\eeq

Next 
we consider the coupling of a photon with the general $q\bar{q}$ mesons. 
By applying the conventional minimal substitution method to 
the free action ({\ref{aiu}}), 
\beq
\partial_{q,\mu}\to \partial_{q,\mu}-iQeA(x_{q})_{\mu}
\eeq
($Q=$diag$(2/3,-1/3,-1/3)$ being quark charge matrix), 
we obtain the action for the relevant EM interaction, 
up to lowest order of coupling constant $e$, 
\beq
{S}^{EM}_{I}&=&\int d^{4}x_{q} d^{4}x_{\bar{q}} 
\sum_{i} {j}_{i,\mu}(x_{q},x_{\bar{q}})A_{\mu}(x_{i}) \nonumber\\
&=& \int d^{4} X {J}_{\mu}(X)A_{\mu}(X)
\ \ \ (i=q, \bar{q}). 
\label{sem}
\eeq
Writing only the terms related to our relevant transition process, 
(\ref{sem}) yields 
\begin{eqnarray}
{j}_{i,\mu}(x_{q},x_{\bar{q}})
&=& -ie \frac{m_{q}+m_{\bar{q}}}{m_{i}}\langle 
\bar{\Phi}
(x,X)Q (
ig_{M}^{(i)}\sigma_{\mu\nu}^{(i)} 
({\overrightarrow{\partial}}_{i,\nu}
+{\overleftarrow{\partial}}_{i,\nu}))\Phi (x,X) \rangle_{S,F}.
\end{eqnarray}
Here we have introduced the parameter $g_{M}^{(i)}$ concerning intrinsic 
magnetic moment of their constituents. 
Integrating on the relative space-time coordinate for the GS-WF, 
we obtain the 
effective spin current of meson in momentum
representation, 
\begin{eqnarray}
&&{J}_{\mu}(P,P^{'})={J}_{q,\mu}(P,P^{'})
+{J}_{\bar{q},\mu}(P,P^{'}),\\
{J}_{q,\mu}(P,P^{'})&=&e\langle 
{\bar{W}^{(-)}}(P^{'})Q[
\frac{m_{q}+m_{\bar{q}}}{m_{q}}g_{M}^{(q)}
i\sigma_{\mu\nu}q_{\nu}]W^{(+)}(P)\rangle_{S,F}~, \label{eq:beforecurrent1}
\\
{J}_{\bar{q},\mu}(P,P^{'})&=&e\langle 
W^{(+)}(P) Q [
\frac{m_{q}+m_{\bar{q}}}{m_{\bar{q}}}g_{M}^{(\bar{q})}i\sigma_{\mu\nu}q_{\nu}] 
{\bar{W}^{(-)}}(P^{'})
\rangle_{S,F}~ ,
\label{eq:beforecurrent2}
\end{eqnarray}
where $q_{\mu}\equiv P_{\mu}- P^{'}_{\mu}$ is the four-momentum of photon, 
$M (M')$ represents the mass of initial (final) meson. Here 
we used the fact referred in the footnote below Eq.(\ref{all}). 
The above framework to treat the EM interaction, 
leading to conserved EM current of hadrons, 
is the same as in our previous 
$\widetilde{U}(12)$-scheme, COQM.{\cite{oldCOQM}} \\

\bfl
({\it Intrinsic electric dipole transition})
\efl

In our scheme, the relativistic covariance of the spin current 
play an important role in some radiative transition processes, 
due to the inclusion of Dirac spinor with negative $\rho_{3}$-value. 
To clarify this point, we rewrite the spin current vertex operator as 
\beq
\sigma_{\mu\nu} iq_{\nu}  A_{\mu}=\sigma_{\mu\nu} F_{\mu\nu} =
{\mbox{\boldmath{$\sigma$}}} \cdot {\bf{B}} - i\rho_{1} 
{\mbox{\boldmath$\sigma$}}\cdot{\bf{E}}~~.
\eeq
In the cases of transition between both positive 
(negative) $\rho_{3}$ Dirac spinors, 
as is well known, a principal contribution comes from 
the magnetic interaction, 
\beq
\bar{u}_{\rho_{3}=\pm}(\mbox{\boldmath $q$})\left( \sigma_{\mu\nu} iq_{\nu} A_{\mu} \right)
u_{\rho_{3}=\pm}(\mbox{\boldmath $0$}) \approx
\chi^{\dagger} \left( \pm {\mbox{\boldmath{$\sigma$}}} \cdot {\bf{B}} \right)
\chi ~~~~~{\rm for} ~~|\mbox{\boldmath $q$}|\approx 
0, 
\eeq
where $\chi$ represents two component $SU(2)_{\sigma}$-spinor. 
In this case, a contribution of the electric interaction, coming from the 
$\sigma_{i4}iq_{i} A_{4}$-term, is appeared as a relativistic 
correction caused by the recoil effect. 
On the other hand, in transitions between Dirac spinors 
with positive and negative $\rho_{3}$-values, 
the electric interaction becomes a principal contribution, 
\beq
\bar{u}_{\rho_{3}=\pm}(\mbox{\boldmath $q$})\left( \sigma_{\mu\nu} iq_{\nu} A_{\mu} \right)
u_{\rho_{3}=\mp}(\mbox{\boldmath $0$}) \approx
\chi^{\dagger}  
\left( \mp i
{\mbox{\boldmath$\sigma$}}\cdot{\bf{E}} \right)
\chi^{}  ~~~~~{\rm for} ~~|\mbox{\boldmath $q$}|\approx 
0. 
\eeq
Accordingly, this ``intrinsic electric dipole''{\cite{Dshad03}} transition 
describes the transition between GS mesons 
accompanied by their parity change. \\

\bfl
({\it Chirality conservation in radiative transition process} )
\efl

By using the current stated above, a very useful selection rule is realized 
in the radiative transition processes. 
That is, the chirality eigen value of the relevant meson
is conserved. 
For an example, 
\beq
&& A^{(E)}(\chi=-) \to P(\chi=-) ~~\gamma ~~~~{\rm ;allowed}\\
&& A^{(N)}(\chi=+)\to P(\chi=-) ~~\gamma ~~~~{\rm ;forbidden}\\
&& V(\chi=+ \oplus - ) \to P(\chi=-) ~~\gamma ~~~~{\rm ;allowed}. 
\eeq

\section{Numerical Results} 

\bfl
({\it Fixing the parameters})
\efl

From the expression of currents, we can derive 
the relevant formulas of radiative decay width, which are given 
in the following Tables {\ref{tab:current}} and {\ref{tab:width}}.

In this work, we take the following values of parameters in our scheme. \\
\begin{itemize}
\item $M,M^{'}$: {take the physical meson mass (PDG value{\cite{PDG2004}}), 
\\~~~~~~~~~~~~or the predicted values (see, K. Yamada's talk{\cite{Yamada}})}, 
\item $g_{M}^{(n)}=1.196$ (determined from 
the $\rho^{\pm} \to \pi^{\pm} \gamma$), 
\item The ratio of constituent quark mass: $x \equiv \frac{m_{n}}{m_{s}}=0.82, $
$g_{M}^{(s)}=1.184 $ (determined from 
the $K^{*\pm}(K^{*0})\to K^{\pm}(K^{0}) \gamma$),
\item flavor mixing angle: 
$\phi_{P}=-45.3^{\circ}$, $\phi_{V}=3.4^{\circ}$, $\phi_{others}=0^{\circ}$. \\
\end{itemize}

\begin{table}[h]
\caption{Effective transition currents for radiative decays 
between the GS mesons. 
Here, $P \equiv P_{s}^{(N)}
$, $S \equiv S^{(N)}$, $V \equiv (V^{(N)} + V^{(E)})/{\sqrt{2}}
$, $V^{'} \equiv (V^{(N)} - V^{(E)})/{\sqrt{2}}
$, and $d \equiv 2(m_{q}+m_{\bar{q}})$. }
\begin{center}
\begin{tabular}{lcc}
\hline
\hline
{process}  
&{current}
&{coupling parameter}\\
\hline
$V \to P \gamma$
&$J_{\mu}=ie\mu_{o}\epsilon_{\mu\nu\rho\alpha}q_{\nu}
\epsilon_{\rho}^{V}P_{\alpha}$ 
&$\mu_{0}=(\frac{d}{2m_{q}}g_{M}^{(q)}\langle P Q V
\rangle+\frac{d}{2m_{\bar{q}}}g_{M}^{(\bar{q})}\langle P V Q\rangle)$ 
$(\frac{1}{\sqrt{2} M})$\\
$V^{'} \to P \gamma$
&$J_{\mu}=ie\mu^{'}_{o}\epsilon_{\mu\nu\rho\alpha}q_{\nu}
\epsilon_{\rho}^{V^{'}}P_{\alpha}$ 
&$\mu_{0}^{'}=(\frac{d}{2m_{q}}g_{M}^{(q)}\langle P Q V^{'}
\rangle+\frac{d}{2m_{\bar{q}}}g_{M}^{(\bar{q})}\langle P V^{'} Q\rangle)$ 
$(\frac{1}{\sqrt{2} M})$\\
$A^{(N)} \to P \gamma$
& $J_{\mu}=0$ (forbidden) 
&$-$\\
$A^{(E)} \to P \gamma$
&$J_{\mu}=ie\epsilon_{A^{(E)}}\epsilon_{\mu}^{A^{(E)}}$
&$\epsilon_{A^{(E)}}=(\frac{d}{2m_{q}}g_{M}^{(q)}
\langle P Q A^{(E)}\rangle+\frac{d}{2m_{\bar{q}}}
g_{M}^{(\bar{q})}\langle P A^{(E)} Q\rangle)(-q v)$\\
$S \to V \gamma$
&$J_{\mu}=-e\zeta \tilde{\epsilon}^{V}_{\mu}$
&$\zeta=(\frac{d}{2m_{q}}g_{M}^{(q)}
\langle V Q S\rangle+\frac{d}{2m_{\bar{q}}}g_{M}^{(\bar{q})}
\langle V S Q\rangle)(q v^{'})\frac{1}{\sqrt{2}}$\\
$A^{(N)} \to V \gamma$
&$J_{\mu}=ie\xi_{0}\epsilon_{\mu\nu\rho\sigma}
q_{\nu}\epsilon_{\rho}^{A^{(N)}}\tilde{\epsilon}^{V}_{\sigma}$
&${\xi}_{0}=(\frac{d}{2m_{q}}g_{M}^{(q)}
\langle V Q A^{(N)}\rangle+\frac{d}{2m_{\bar{q}}}g_{M}^{(\bar{q})}
\langle V A^{(N)} Q\rangle)\frac{1}{\sqrt{2}}$\\
\hline
\end{tabular}
\label{tab:current}
\end{center}
\end{table}
\begin{table}[h]
\caption{Formulae for radiative dacy width. Here, the $\omega$ and $\omega_{3}$ 
are defined as 
$\omega\equiv -v \cdot v^{'}$ and $\omega_{3}\equiv \sqrt{\omega^{2}-1}$, respectively. 
The $\alpha$ denotes the fine structure constant, $\alpha=1/137$. 
}
\begin{center}
\begin{tabular}{lc}
\hline
\hline
{process}
& {$\Gamma$}\\
\hline
$V (V^{'}) \to P \gamma$ \ \ \ \ \ 
& $\Gamma=\frac{\alpha}{3}\mu_{0}^{(')2} |\mbox{\boldmath $q$}|^{3}$\\
$A^{(E)} \to P \gamma$ \ \ \ \ \  
& $\Gamma=\frac{\alpha}{3}{\epsilon_{A^{(E)}}}^2 \frac{|\mbox{\boldmath $q$}|}{M^2}$  \\
$S \to V \gamma$ \ \ \ \ \ 
&$\Gamma=\frac{\alpha}{2J+1}{\zeta}^2 \frac{|\mbox{\boldmath $q$}|}{M^2}$ \\
$A^{(N)} \to V \gamma$ \ \ \ \ \ 
&$\Gamma=\frac{2\alpha}{3}\frac{{\xi_{0}}^2}{M^2}\omega(\omega+\omega_{3}){|\mbox{\boldmath $q$}|}^3$ \\
\hline
\end{tabular}
\label{tab:width}
\end{center}
\end{table}


\bfl 
({\it 1. Radiative decays of $V$-meson})
\efl

As was discussed in \S 2, we use the $V$- and $P$-type BW functions, Eq.({\ref{Vec}}) 
and Eq.({\ref{P}}), as 
the spin WFs of the physical $\rho$-nonets and $\pi$-nonets, respectively. 
A possible decay mode of $V$-meson is only $P\gamma$, as well as conventional scheme. 
The results in comparison with experiments are shown in Table \ref{tab:width-a}. 
Our estimated widths seems to be consistent 
with the experimental data. Here we emphasize that 
it comes from our choice of the $SU(2)_{\rho}$ WF. 
the $\pi$-on spin WF as $i\gamma_{5}/2$ and 
the $\rho$-meson spin WF $V$ as pure Paulon with 
equal weight superposition of $V^{(N)}$ and $V^{(E)}$. \\
\begin{table}[t]
\caption{Predicted radiative decay widths of pure Pauli- vector meson 
$V\to P\gamma$ process in 
comparison with experiments. The input values are underlined. }
\begin{center}
\begin{tabular}{lcc}
\hline
\hline
{$V \to P \gamma $ process} \ \ \ \ \ \ &
{$\Gamma_{\rm theor}$(keV)} \ \ \ \ \ \ 
& {$\Gamma_{\rm exp}$(keV)} \ \ \  \ \ \  \\
\hline
$\rho^{\pm}\to \pi^{\pm} \gamma$ & $\underline{68}$ & $68 \pm 7 $\\
$\rho^{0}\to \eta^{} \gamma$ & $45.2$ & $45.09 $\\

$K^{*\pm}\to K^{\pm} \gamma$ & $\underline{50}$ & $ 50 \pm 5 $\\
$K^{*0}\to K^{0} \gamma$ & $\underline{116}$ & $116\pm 10 $\\

$\omega \to \pi^{0} \gamma$& $620$ & $757.3$\\
$\omega \to \eta^{} \gamma$& $4.20$ & $4.16$\\

$\phi^{}\to \pi^{0} \gamma$ & $2.96$ & $5.24$\\
$\phi^{}\to \eta \gamma$ & $70.7$ & $55.2$\\
$\phi^{}\to \eta^{'} \gamma$ & $0.327$ & $0.264$\\

$\eta^{'} \to \rho^{0} \gamma$ & $72.7$ & $59.59$\\
$\eta^{'} \to \omega \gamma$ & $9.06$ & $6.12$\\
\hline
\end{tabular}
\end{center}
\label{tab:width-a}
\end{table}

\bfl 
({\it 2. Radiative decays of $V^{'}$-meson} )
\efl

In our scheme, the existence of ``extra-vector''  $V^{'}$-nonet 
as ``pure chiral state'' is required as the $q\bar{q}$ GS. 
The plausible candidates of extra-vector meson 
are pointed out in nearly 1.3 GeV region. 
(For more detail, see, K. Yamada{\cite{Yamada}} and 
T. Komada{\cite{Komada}}'s talks.) 
Possible decay modes of $V^{'}$-meson are
(a) $V^{'} \to P \gamma$, 
(b) $V^{'} \to S^{(N)} \gamma$, 
(c) $V^{'} \to S^{(E)} \gamma$, 
(d) $V^{'} \to V \gamma$ , 
(e) $V^{'} \to A^{(N)} \gamma$, 
(f) $V^{'} \to A^{(E)} \gamma$. 
Here we show the result only for the cace of (a) in Table {\ref{tab:width-b}}. 
There is few experimental data be compared with
these estimated numerical values. 
Taking naively the well-known $K^{*}(1410)$ meson
as a $I=1/2$ member of $V^{'}$-nonet, then it seems that 
the predicted width is much larger than experimental one. 
(The only one reported experimental value{\cite{KTeV}} 
for the relevant process is $\Gamma_{\rm exp}<52.9$keV). 
This problem may be possibly resolved by considering the mixing effect of 
the GS $V^{'}$ nonet with ${}^{3}P_{1}/{}^{1}P_{1}$ vector nonet in our scheme.\\
\begin{table}[t]
\caption{Predicted radiative decay widths of ``extra-'' pure chiral- vector meson 
$V^{'} \to P\gamma$ process. }
\begin{center}
\begin{tabular}{lcc}
\hline
\hline
{$V^{'} \to P \gamma $ process} \ \ \ \ \ \ &
{$\Gamma_{\rm theor}$(keV)} \ \ \ \ \ \ 
& {$\Gamma_{\rm exp}$(keV)} \ \ \ \ \ \  \\
\hline
$\rho^{'\pm}(1290)\to \pi^{\pm} \gamma$ & $120$ & $$\\
$\rho^{'0}(1290)\to \eta^{} \gamma$ & $329$ & $ $\\
$\rho^{'0}(1290) \to \eta^{'} \gamma$ & $47.6$ & \\

$K^{*'\pm}(1410)\to K^{\pm} \gamma$ & $161$ & $  $\\
$K^{*'0}(1410)\to K^{0} \gamma$ & $372$ & $ < 52.9${\cite{KTeV}}\\

$\omega^{'}(1290) \to \pi^{0} \gamma$& $894$ & \\
$\omega^{'}(1290) \to \eta^{} \gamma$& $36.5$ & \\

$\phi^{'}(1540) \to \pi^{0} \gamma$ & $0$ & \\
$\phi^{'}(1540)\to \eta\gamma$ & $186$ & \\
$\phi^{'}(1540)\to \eta^{'} \gamma$ & $73.0$ & \\

$\omega^{'}(1290) \to \eta^{'} \gamma$ & $5.28$ & \\
\hline
\end{tabular}
\label{tab:width-b}
\end{center}
\end{table}

\bfl 
({\it 3. Radiative decays of $S^{(N)}$-meson})
\efl

One of most important chiral multiplet 
in the $q\bar{q}$ GS is the scalar $\sigma$-nonet. All members 
have now established experimentally.
Possible decay modes of $S^{(N)}$-meson is only $V \gamma$, 
as well as in the conventional scheme. 
The results in comparison with experiments and with the other models 
are shown in Table \ref{tab:width-c}. Note that it is not considered 
the finite width effect of $\kappa$- and $\sigma$-mesons and 
the coupled channel ($K^{+}K^{-}$-loop) effect 
which may play essential role in this process{\cite{Achasov}} 
in this calculation. \\
\begin{table}[t]
\caption{Predicted radiative decay widths of 
$S^{(N)} \to V\gamma$ process in comparison with other model and experiments. }
\begin{center}
\begin{tabular}{lcrrr}
\hline
\hline
{$S^{(N)} \to V \gamma $ process} \ \ \ \ \ \ &
{$\Gamma_{\rm theor}$(keV)} 
& {$\Gamma_{\rm exp}$(keV)} &
{$\Gamma_{{}^{3}P_{0}}$(keV)} 
& {$\Gamma_{4q}$(keV)} \\
&&&in NRQM{\cite{Kalash}}
&with VMD model{\cite{Harada}} \\
\hline
$a_{0}(980)^{\pm}\to \rho^{\pm} \gamma$ & $25.1$ &$-$ & $14$ & $3.0$ \\
$a_{0}(980)^{0} \to \omega \gamma$ & $203$ &$-$ & $125$ & $641$\\

$\kappa_{A}^{\pm}(1105) \to K^{*\pm}(892) \gamma$ & $36.5$ & $-$ & $-$ & $-$\\
$\kappa_{A}^{0}(1105)\to K^{*0}(896)  \gamma$ & $79.1$ & $-$ & $-$ & $-$\\

$\rho^{0} \to \sigma(600) \gamma$& $73.3$ &$-$ &$-$ & $0.23/17$ \\
$\omega \to \sigma \gamma$& $8.99$ & $-$& $-$ & $16/33$ \\
$\phi^{}\to \sigma \gamma$ & $0.280$ & $-$ & $-$ & 137/33 \\

$f_{0}(980) \to \rho^{0} \gamma$ & $0$ & $-$ & $0$& $19/3.3$\\
$f_{0}(980) \to \omega \gamma $ & $0.292$ &$-$ & $0.109$ & $126/88$\\

$\phi\to f_{0}(980) \gamma$ & $0.182$ &$1.87\pm0.11$ & $0.18$ & $-$\\
$\phi\to a_{0}(980) \gamma$ & $0.000987$ & $0.32\pm0.026$& $0.0013$ & $-$ \\
\hline
\end{tabular}
\label{tab:width-c}
\end{center}
\end{table}

\bfl
({\it 4. Radiative decays of $A^{(E)}$-meson})
\efl

The $1^{+-}$-nonet are classified as the excited ${}^{1}P_{1}$ state 
in the conventional classification scheme. However, 
there is another $1^{+-}$-nonet as $A^{(E)}$-type GS chiralon 
in our scheme. 
Its possible radiative decay mode are 
(a) $A^{(E)} \to P\gamma$, 
(b)$A^{(E)} \to S_{A} \gamma$, 
(c)$A^{(E)} \to V \gamma$. 
Their estimated widths are shown in the Table \ref{tab:width-d} 
only the case of (a). 
\begin{table}[h]
\caption{Predicted radiative decay widths of $A^{(E)} \to P\gamma$ process in 
comparison with experiments. }
\begin{center}
\begin{tabular}{lcc}
\hline
\hline
{$A^{(E)}$ process} \ \ \ \ \ \ &
{$\Gamma_{\rm theor}$(keV)} \ \ \ \ \ \ 
& {$\Gamma_{\rm exp}$(keV)} \ \ \ \ \ \  \\
\hline
$b_{1}(1235)^{\pm}\to \pi^{\pm} \gamma$ & $229$ &$230 \pm 60$ \\
$b_{1}(1235)^{0} \to \eta  \gamma$ & $586$ &$$ \\
$b_{1}(1235)^{0}\to \eta^{'} \gamma$ & $61.0$ &$$ \\
$h_{1}(1170) \to \pi^{0} \gamma$ & $1957$ &$$ \\
$h_{1}(1170) \to \eta    \gamma$ & $57.3$ &$$ \\
$h_{1}(1170) \to \eta^{'} \gamma$ & $3.80$ &$$ \\
$K_{1B}(1350)^{\pm}\to K^{\pm} \gamma$ & $297$ &$$ \\
$K_{1B}(1350)^{0} \to K^{0} \gamma$ & $682$ &$$ \\
$h_{1}(1380) \to \pi^{0} \gamma$ & $0$ &$$ \\
$h_{1}(1380) \to \eta    \gamma$ & $299$ &$$ \\
$h_{1}(1380) \to \eta^{'} \gamma$ & $79.0$ &$$ \\
\hline
\end{tabular}
\label{tab:width-d}
\end{center}
\end{table}

Here the result on the $b_{1}(1235)$-decay is specially to be noted : 
In the previous scheme, 
our predicted width for 
$b_{1}({}^{1}P_{1})\to \pi ({}^{1}S_{0})$ ($\Gamma_{\rm theor}=66$ keV) was much smaller 
than experimental one ($\Gamma_{\rm exp}= 230\pm 60$ keV). 
We had considered at that time that this discrepancy is quite serious for the scheme, 
since to this process contributes only the convection current, 
whose form reflects the conservation of EM current. 
However, it has been shown now that the experiment 
can be reproduced by assigning $b_{1}(1235)$ as 
the GS chiralon. We give the other results concerning the $A^{(E)}\to P \gamma$. \\
{\it Remarks on $K_{1}(1270)$ and $K_{1}(1400)$}:~~
Concerning the radiative decays of $K_{1}(1270)$- and $K_{1}(1400)$-meson, 
the experimental data,{\cite{KTeV}}~ $\Gamma(K_{1}(1270)\to K^{0}\gamma )=
73.2$ keV, $\Gamma(K_{1}(1400)\to \phi \gamma )=280.8$ keV
are reported, respectively. 
We examined this results by assuming these states being mixed states of $K_{1}^{(N)}$ 
and $K_{1}^{(E)}$, finding there is no solution of $(N)$-$(E)$ mixing angle parameter 
consistent with the both above. 
Thus, at least, the one of them is not pure chiral state.\\

\bfl
({\it 5. Radiative decays of $A^{(N)}$-meson})\\
\efl

As shown in the previous section, 
the $A^{(N)}$-nonet, which is the GS 
axial-vector chiral state, 
does not have the decay mode $A^{(N)} \to P\gamma$ due to chirality conservation rule.
This fact indicates that 
well known $a_{1}(1260)$ meson 
decaying to $\pi\gamma$ is not a pure GS chiralon at least. 
This is consistent to our previous result by COQM that $a_{1}(1260)\to \pi \gamma$ 
could be explained as the decay of the ${}^{3}P_{1}$ Pauli-state. 
We predict the another $A^{(N)}$-nonoet, being different from 
the ${}^{3}P_{1}$-state, in the lower mass region compared to $1260$ MeV. 
Possible radiative decay modes are (a) $A^{(N)} \to V\gamma$ and 
(b) $A^{(N)} \to S^{(E)} \gamma$. Their estimated widths 
are shown in Table {\ref{tab:width-e}} only the case of (a). \\
\begin{table}[t]
\caption{Predicted radiative decay widths of $A^{(N)} \to V \gamma$ process. } 
\begin{center}
\begin{tabular}{lc}
\hline
\hline
{$A^{(N)} \to V \gamma $ process} \ \ \ \ \ \ &
{$\Gamma_{\rm theor}$(keV)} \ \ \ \ \ \ 
 \\
\hline
$a_{1}(1210)^{\pm} \to \rho^{\pm} \gamma$ & $82.1$  \\
$a_{1}(1210)^{0} \to \omega \gamma$ & $701$  \\
$a_{1}(1210)^{0} \to \phi \gamma$ & $0.218$  \\
$f_{1}(1420)^{0} \to \omega \gamma$ & $2.74$  \\ 
$f_{1}(1420)^{0} \to \rho^{0} \gamma$ & $0$  \\ 
$f_{1}(1420)^{0} \to \phi \gamma$ & $179$  \\ 
$K_{1A}(1328)^{0} \to K^{*0} \gamma$ & $268$  \\ 
$K_{1A}(1328)^{\pm} \to K^{*\pm} \gamma$ & $120$  \\ 
$f_{1}(1210)^{0} \to \omega \gamma$ & $77.8$ \\
$f_{1}(1210)^{0} \to \phi \gamma$ & $0.024$  \\
$f_{1}(1210)^{0} \to \rho^{0} \gamma$ & $739$  \\
\hline
\end{tabular}
\label{tab:width-e}
\end{center}
\end{table}
{\it Remarks on $f_{1}(1285)$}:~~From the experimental data, we have 
$\Gamma(f_{1}(1285)\to \rho^{0}\gamma )=
1325$ keV\cite{PDG2004} / $674.8$ keV\cite{VES}, 
$\Gamma(f_{1}(1285)\to \phi \gamma )=17.834$ keV.\cite{PDG2004} 
We examined this results by assuming $f_{1}(1285)$ being $f_{1}^{(N)}$ chiralon, 
finding there is no solution of flavor octet-singlet mixing angle parameter 
consistent with the both above. 
Thus, $f_{1}(1285)$ is not pure chiral state as well as $a_{1}(1260)$.


\section{Concluding Remarks}

We have reinvestigated the radiative decay of the GS light quark mesons 
in our new  scheme. As a result, it is shown that; 
(1)~Experimental data on the ordinary $V \to P \gamma$ process
are well reproduced in the $\widetilde{U}(12)$-scheme, as well as our 
previous COQM results. The essential points are that 
to take the $\pi$-on spin WF as $i\gamma_{5}/2$, reflecting its Goldston-boson nature, 
and to take the $\rho$-meson spin WF 
as being a pure-Pauli state. 
(2)~ The $b_{1}(1235)$, has been classified as the 
$P$-wave excitation in the conventional 
classification scheme, is a promising candidate of the chiral states. 
(3)~It is also shown that 
$K_{1}(1270)$ and $K_{1}(1400)$ are not both pure chiral states, 
and also  $K^{*}(1410)$, $f_{1}(1285)$ is not pure chiral state. 
So, we have to consider the possibility of mixing them with the P-wave states.
(4)~The radiative decay widths of $V^{'}$-, $S^{(N)}$-, 
$A^{(N)}$-, and $A^{(E)}$-mesons are predicted. 
They should be checked experimentally.

\section*{Acknowledgements}

We are grateful to Kunio Takamatsu and other member of the sigma group 
for useful discussions. 
%

\end{document}